\documentclass[a4paper,10pt]{article}
\usepackage[utf8]{inputenc}
\usepackage{multirow} 
\usepackage{booktabs} 
\usepackage{amssymb}
\usepackage{amsmath}
\usepackage{setspace}
\usepackage{graphicx}
\usepackage{amsmath}
\usepackage{color}
\usepackage{gensymb}
\usepackage[colorlinks,citecolor=blue,urlcolor=blue,linkcolor=blue]{hyperref}
\usepackage{authblk}

\title{Phaseless Microwave Imaging Of Dielectric Cylinders: An Artificial Neural Networks--Based Approach}
\author[1]{Jes\'us E. Fajardo}
\author[2]{Juli\'an Galv\'an}
\author[3]{Fernando Vericat}
\author[4]{Carlos M. Carlevaro}
\author[5]{Ramiro M. Irastorza*}
\affil[1,3,4,5]{\textit{Instituto de F\'isica de L\'iquidos y Sistemas Biol\'ogicos CONICET - CCT La Plata.}\newline \textit{La Plata, Buenos Aires, Argentina}}
\affil[2]{\textit{Instituto Argentino de Radioastronom\'ia CONICET - CCT La Plata}\newline \textit{La Plata, Buenos Aires, Argentina}}
\affil[4]{\textit{Grupo de Materiales Granulares, Departamento de Ingenier\'ia Mec\'anica, FRLP, Universidad tecnol\'ogica Nacional.}\newline \textit{La Plata, Argentina.}}
\affil[5]{\textit{Instituto de Ingenier\'ia y Agronom\'ia, UNAJ.}\newline \textit{Florencio Varela, Buenos Aires, Argentina}}

\begin{document}

\maketitle





\begin{abstract}

An inverse method for parameters estimation of infinite cylinders (the dielectric properties, location, and radius) in two dimensions from amplitude-only microwave information is presented. To this end two different Artificial Neural Networks (ANN) topologies are compared; Multilayer Perceptron (MLP) and a Convolutional Neural Network (CNN). Several simulations employing the Finite Differences in Time Domain (FDTD) method are performed to solve the direct electromagnetic problem and generate training, validation, and test sets for the ANN models. The magnitude of the mean errors in estimating the position and size of the cylinder are up to (1.9 $\pm$ 3.3) mm and (0.2 $\pm$ 0.8) mm for the MLP and CNN, respectively. The magnitude of the mean percentage relative errors in estimating the dielectric properties of the cylinder are up to (6.5 $\pm$ 13.8) \% and (0.0 $\pm$ 7.2) \% for the MLP and CNN, respectively. The errors in the parameters estimation from the MLP model are low, however, significantly lower errors were obtained with the CNN model. A validation example is shown using a simulation in three dimensions. Measurement examples with homogeneous and heterogeneous cylinders are presented aiming to prove the feasibility of the described method. 

\end{abstract}

%
%

\section{Introduction}
\label{sec:INTRO}
Inverse problem in microwave imaging from phaseless data (amplitude--only) is a challenging problem \cite{LibroPastornino10}. It gained attention because of the simpler measurement apparatuses requirements and its consequent costs reduction. To solve this problem two main approaches were proposed: using phase retrieval methods \cite{costanzo2015hybrid}, and reconstructing the scatterers directly from measurements \cite{caorsi2003electromagnetic}. In the former, the phase is estimated from the amplitude--only data, and then amplitude and phase are used with the traditional reconstruction algorithms. In the latter, several approches were developed using deterministic \cite{li2008tomographic,li2008two} and stochastic methods \cite{bermani2002microwave,alvarez2016inverse}. As a stochastic method, Artificial Neural Networks (ANN) have advantageous capabilities to account for non--linear effects present in inverse scattering problems \cite{li2018deepnis}. A variety of inverse scattering ANN--based approaches were developed, including hybrid methods combining ANN and iterative methods \cite{kamilov2016recursive}. Particularly for phaseless scenarios, one of the first works in this direction was proposed in the earliest 2000's by Bermani et al. \cite{bermani2002microwave}. The authors employed a Multilayer Perceptron (MLP) to solve (not simultaneously) two problems: reconstructing the dielectric properties of a cylinder, with known radius and postition; and detecting the position of a buried cylinder, with known dielectric properties and radius. 

In the field of Deep Learning, the improvements which provide Convolutional Neural Networks (CNN's) respect to MLP's for recognizing patterns in images and complex data are notorious \cite{goodfellow2016deep}. Recently, some works presented different CNN--based approaches for solving the inverse scattering problem, showing the potential of the model \cite{wei2018deep,li2018deepnis,jin2017deep}. The CNN's of these works are based on the U--Net \cite{ronneberger2015u} and use both, amplitude and phase information. Surprisingly, to our knowledge, the CNN for solving an Electromagnetic (EM) inverse problem with amplitude--only information, has not been employed yet. In this work we implemented such a topology and compared the results with a MLP. We computationally addressed the inverse scattering problem of estimating simultaneously dielectric and geometric parameters of an infinite cylinder. The homogeneous cylinder was illuminated by a circular array of monopole antennas (similar to the array presented by Meaney et al. \cite{meaney2012b}) and the electric field magnitude was computed by finite differece time--domain (FDTD) in a two dimensional (2D) simplified model. Such simulations were used to train the ANN. Using the trained ANN heterogeneous cylinders were evaluated aswell and the predicted effective parameters are commented. In order to show the validity of the method, a full three dimensions simulation (using Finite Element Method (FEM)) is also provided, in which the whole setup is modeled as realistic as possible. 

\section{Materials and Methods}
\label{sec:Materiales}
\subsection{Direct EM Scattering Problem Solution}
The EM direct problem was solved using FDTD, implemented with the free available software MEEP \cite{paperMeep10}. The transmitter monopole antenna was considered as a line of current (a point source in 2D) emitting a TM-polarized electric field $E_{\text{z}} \propto e^{j\omega t}$, being ``$z$'' the axis parallel to that of the antennas, $\omega=2\pi f$ and $f$ is the frequency of 1.1 GHz. The receiver antenna was not modeled,  the value of the electric field at the nearest point in the grid was collected instead. The geometry is a simplification of the one presented by Meaney et al. \cite{meaney2012b}. It consists in a circular array of 16 monopole antennas equally angularly spaced and disposed in a circle with a diameter of 15.2 cm. The object under study (e. g.: homogeneous cylinder) was placed within the investigation domain, which is a concentric circle with a diameter of 14.0 cm. The coupling bath is a glycerin-water mixture in 80:20 proportion with dielectric properties $\varepsilon_{r} = 28.6$ and $\sigma = 1.26$ Sm$^{-1}$ at 1.1 GHz. The size of the simulation box is 25 cm $\times$ 25 cm (with the array in the center), the spatial grid resolution of the simulation box was 1.0 mm and the Courant factor was 0.5. The boundary conditions were Perfectly Matched Layers (PML), i.e., total absorption in the box edges.\\
In order to validate the FDTD models, the simulations were compared to analytical known results (not shown here) \cite{harrington1961time,arslanagic2006electric}. A total of 10000 simulations of the direct EM problem were performed with different scatterer cylinders. The location, radius and dielectric properties of the cylinders were randomly varied (uniform distributions) within the intervals described in Tab. \ref{TableI}. From these simulations the total field magnitude ($|E_{\text{t}}|$) was obtained. A simulation for the incident magnitude field ($\left|E_{\text{inc}}\right|$) was also computed (without the scatterer cylinder) for normalization purposes. The simulation of each particular cylinder implies that 16 (number of transmitters) $\times$ 16 (number of receivers excluding the transmitter) = 256 values were saved. 

\begin{table}[h!]
\centering
\caption{Simulated range of the dielectric properties (at 1.1 GHz) and radius of the cylinder.}
\label{TableI}
\begin{tabular}{lccc} 
\hline
Radius (mm)& $\varepsilon_{r}$ & $\sigma$ (Sm$^{-1}$) \\
\hline
[2.0, 30.0] & [10.0, 80.0] & [0.40, 1.60] \\
\hline
\end{tabular}
\end{table}

A whole setup simulation was also computed in 3D for validation purposes. Simulations were performed using \textit{ad hoc} software applying the FEM approach (similar to the work of Attardo et al. \cite{attardo2012whole}). The model is composed of: two monopole antennas (transmitter and receiver, which are modeled by a coaxial cable with the inner connector extended $\lambda/4$), the cylinder (with fixed postition in the center of the array), and the acrylic tank container. The simulation box is a cube of 30 cm of side. A total of eigth (one transmitter and eigth receivers) 3D simulations were computed. The 256 data needed were obtained by the simetry of the model (the cylinder is in the center).
\subsection{Inverse problem using ANN}

The inverse problem in microwave tomography is defined as reconstructing the object under study (the map of its dielectric properties which we called \textit{the outputs}) having available the data (\textit{the inputs}) in the observation points (the circular array of 16 antennas). The inputs ($x_{i}$ with $i = 1,\ldots,256$) of the network corresponds to the ratio $|E_\text{t}|/|E_{\text{inc}}|$ normalized to the $\left[0,1\right]$ interval (16$\times$16 = 256 values). The outputs ($y_{i}$ with $i=1,\ldots,5$) of the model are: the coordinates of the cylinder center ($X_{\text{cen}},Y_{\text{cen}}$), its radius, and the dielectric properties of the cylinder at 1.1 GHz ($\varepsilon_r, \sigma$). In this work, a simplified version of this problem is addressed by using ANN. It is basically a problem of minimization of the difference between the output of the model and the actual output.

The two kind of networks commented below were implemented using the Application Programming Interface Keras \cite{keras} with Tensorflow \cite{tensorflow} as backend package.

\subsubsection{MLP Overview and Topology}
A MLP ANN is an interconnected network of artificial neuron-like units, disposed in layers, where all the units of each layer are connected with all the units of the previous and next
layer, but there are not connections between the units in the same layer. In this kind of ANN, the output of each layer is given by:
\begin{equation}
\mathbf{x'}=\mathbf{f(Wx+b)}
\label{mlp}
\end{equation}
where $\mathbf{f}$ is the activation function of the layer, $\mathbf{W}$ is a matrix of adjustable parameters and $\mathbf{x}$ and $\mathbf{b}$ are, respectively, the input vector of the layer and a vector containing the bias term of each neuron of that layer. In general, in a network composed by $N$ layers, the whole operation is a nested composition of these operations, given by:
\begin{equation}
\tilde{\mathbf{y}}=\mathbf{f_N(W_Nf_{N-1}(\cdots (W_2f_1(W_1x+b_1)+b_2)\cdots +b_{N-1})+b_N)}
\label{mlp2}.
\end{equation}
The input vector $\mathbf{x}$ is propagated through the network and the output vector $\tilde{\mathbf{y}}$ of parameters are estimated, then the result is compared to the actual $\mathbf{y}$ vector and a loss is computed. Tipically, one of the loss metrics used in such a regression is the Mean Absolute Percentage Error (MAPE), if the the number of parameters is $n$ (size of the vector $\mathbf{y}$) then:
\begin{equation}
\text{MAPE}=\frac{1}{n}\sum_{i=1}^n\left|\delta_{i} \right|
\label{mae}
\end{equation}
where
\begin{equation}
 \delta_{i} = \frac{\tilde{y}_{i}-y_i}{y_i}
 \label{delta}
\end{equation}
and $\tilde{y_i}$ and $y_i$ are the output of the model and the actual value of this $i$--th parameter (components of the vectors $\tilde{\mathbf{y}}$ and $\mathbf{y}$), respectively. In this work we also computed the Mean Absolute Error (MAE) which is directly computed as: 
\begin{equation}
\text{MAE}=\frac{1}{n}\sum_{i=1}^n\left|\tilde{y}_{i}-y_i\right|. \label{mae2}\end{equation}
After evaluating the loss, the gradients are retropropagated through the network from $\mathbf{y}$ to  $\mathbf{x}$ and the weights are updated in all the neuron-like units composing the network
in a direction opposite to that of the gradients $\nabla_{\theta}J(\theta)$, where $J(\theta)$ is the objective or loss function parameterized to the model parameters $\theta$. Several approaches are used for doing this task \cite{GDoverview}. Once the validation loss is low enough and the network is capable of generalizing the prediction for new input vectors (not used during the training period), the matrices with the weights are saved and its performance is evaluated in the test set.\\

The implemented topology is described in Table \ref{TableII}. A total of 6000 simulations were used for training, 2000 simulations for the validation set, in order of preventing the model from overfitting, and 2000 simulations were reserved for test purposes. The input of the network corresponds to the ratio $|E_\text{t}|/|E_{\text{inc}}|$ normalized to the $\left[0,1\right]$ interval. The output vector of the model is composed by 5 parameters: the coordinates of the cylinder center ($X_{\text{cen}},Y_{\text{cen}}$), its radius and dielectric properties ($\varepsilon_r, \sigma$). As activation functions, Rectifier linear units (ReLU \cite{nair2010rectified}) functions were used in all the neurons-like units, except in the output layer, where a general linear function was used. During the network training, a variation of the standard Gradient Descent (GD) method with adaptative momentum (presented by Kingma and Ba \cite{adam}) was used, which, for this particular study, outperformed the traditional ones (like those used in \cite{GDoverview}). The selected loss function was MAPE (Eq. \ref{mae}) and 1000 epochs were used for training the network with a batch size of 50.\\

\begin{table}[h!]
\centering
\caption{\footnotesize Topology of the implemented MLP.}
\label{TableII}
\begin{tabular}{lccc} 
\hline
Input Layer& Hidden Layers & Output Layer \\
\hline
256 units & 3 $\times$ 100 units & 5 units \\
 & Activation: ReLU & Activation: Linear \\
\hline
\end{tabular}
\end{table}

\subsubsection{CNN Overview and Topology}
CNN can also be considered a nested composition of functions, from a $d$-dimensional input space $\mathbf{X} \in \mathbb{R}^d$ to $\mathbf{y} \in \mathbb{R}^n$, which is the output space of the inferred variables. The difference respect to the MLP lies in the fashion that the internal operations are made, which allows the CNN to extract local features at different complexity levels. 
Particularly, the operation described by Eq. (\ref{mlp}) changes into  convolutional ones, which are the building blocks of CNNs and are defined on a translation invariance basis and have shared weights across different spatial locations. Both the input and the output of convolutional layers are 2D matrices called feature maps, where the output feature map is obtained by convolving convolution kernels on the input feature map as:
\begin{equation}
\tilde{\mathbf{y}}=\mathbf{f}_{\text{s}}(\mathbf{X;W,b})=\mathbf{W}*_\text{s}\mathbf{X+b}
\end{equation}
Here $\mathbf{X}$ is the input feature map; $\mathbf{W}$ and $\mathbf{b}$ denote kernel and bias, respectively; $*_\text{s}$ represents convolution operation with stride $\text{s}$. As a result, the resolution of the output feature map $\mathbf{f}_{\text{s}}(\mathbf{X;W,b})$ is downsampled by a factor of $\text{s}$. The output feature map of the last convolutional layer can then be fed into a stack of fully connected layers (MLP), which discard the spatial coordinates of the input and generates a global estimation for the input image--like matrix \cite{goodfellow2016deep}.\\

\begin{figure}[h!]
\centering
\includegraphics[width=0.5\columnwidth]{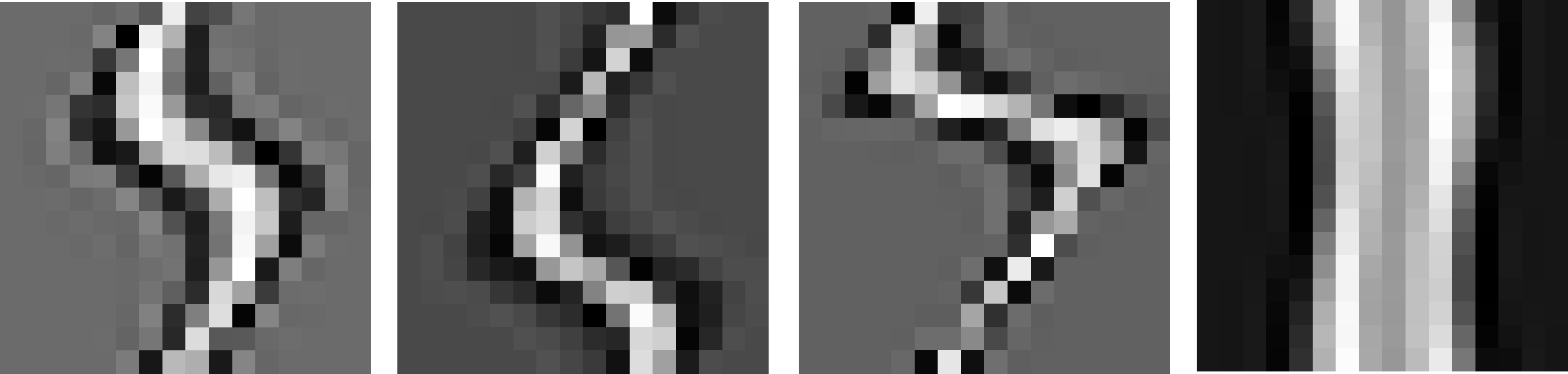}
\caption{Input examples: black pixels corresponding to lower values and the white ones to higher values of the $|E_\text{t}|/|E_{\text{inc}}|$ ratio.}\label{Fig1}
\end{figure}

A detailed architecture of the implemented CNN is shown in Table \ref{TableIII}. The sizes of the sets, number of epochs, the batch size, and GD method were the same to those used by the MLP. The input dimension is also the same input information used for the MLP, but reshaped to 16 $\times$ 16 image--like matrices (Fig. \ref{Fig1}).

\begin{table}[h!]
\centering
\caption{\footnotesize Topology of the implemented CNN.}
\label{TableIII}
\setlength\tabcolsep{1.5pt}
\begin{tabular}{lccccc} 
\hline
Layer & Activation & Kernel & Filters & Input & Output \\
\hline
2D Conv. & ReLU & 3 $\times$ 3 & 32 & (16, 16, 1) & (14, 14, 32) \\
 & & Stride: 1 &  &  &  \\
\hline
2D Conv. & ReLU & 3 $\times$ 3 & 64 & (14, 14, 32) & (12, 12, 64) \\
 & & Stride: 1 &  &  &  \\
\hline
Pool & ReLU & 2 $\times$ 2 & 64 & (12, 12, 64) & (6, 6, 64) \\
\hline
Dense & ReLU & - & - & 2304 & 100 \\
Dense & ReLU & - & - & 100 & 100 \\
Dense & Linear & - & - & 100 & 5 \\
\hline
\end{tabular}
\end{table}

\section{Results and discussion}
\label{sec:Resultados}

\subsection{Comparison MLP and CNN}
\begin{figure}[h]
\centering
\includegraphics[width=0.8\columnwidth]{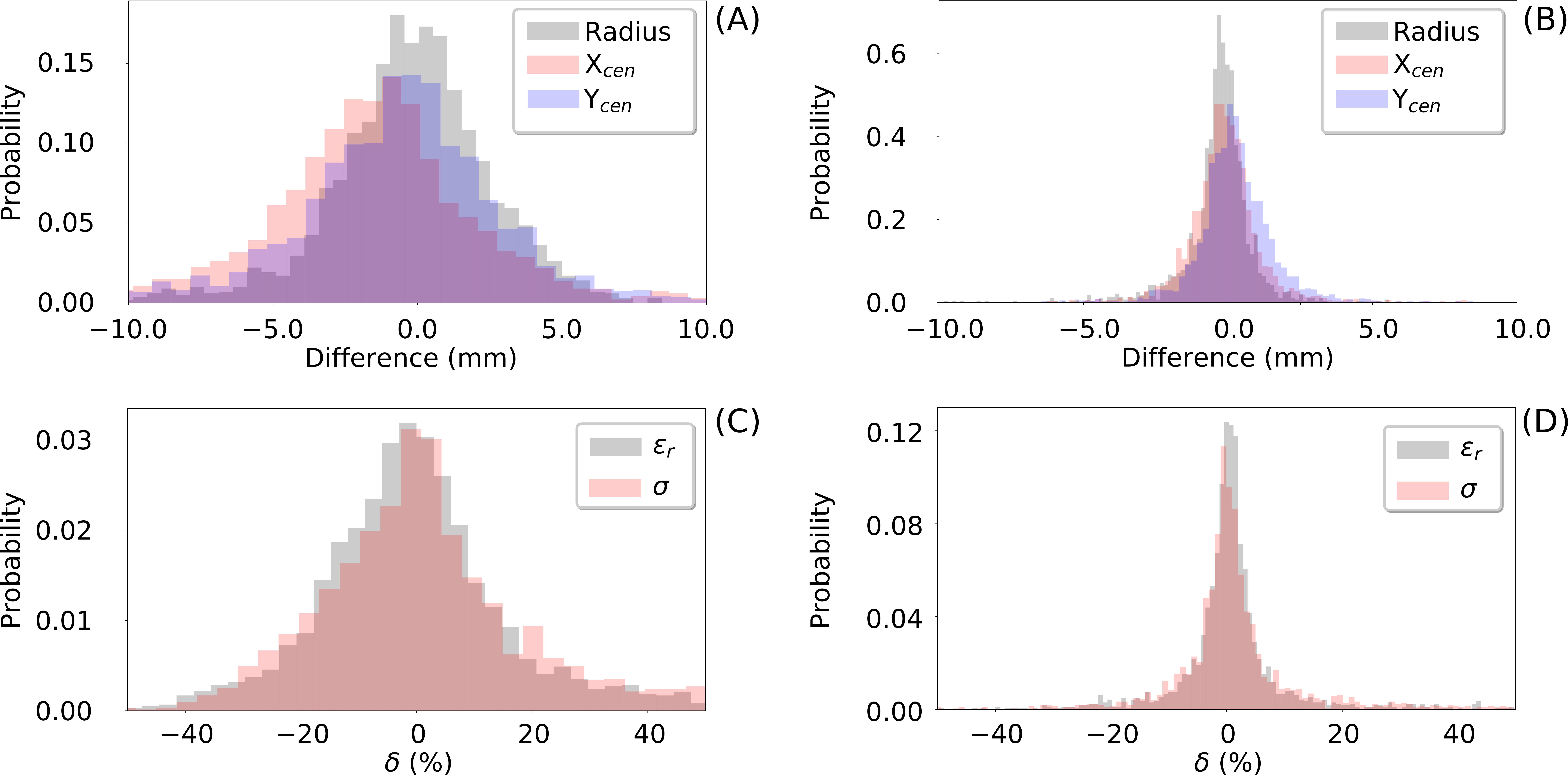}
\caption{Histograms of geometric (A) and dielectric (C) estimation errors using the MLP model. Histograms of geometric (B) and dielectric (D) estimation errors using the CNN model.}\label{Fig2}
\end{figure}

We proposed an ANN similar to that presented by Bermani et al. \cite{bermani2002microwave} but in our work the estimations are obtained simultaneusly (dielectric and geometric properties). Comparison of the estimation quality for each parameter using both, the MLP and CNN models are shown in Fig. \ref{Fig2}. The histograms show the percent error of the dielectric parameters and the error in the geometric parameters. A good behaviour is observed for both MLP Networks and CNNs. Distributions of errors are zero centered and symmetric bell--shaped. CNNs show considerably narrower distributions of the errors as compared to MLP for both dielectric and geometric parameters.\\

\begin{figure}[h]
\centering
\includegraphics[width=1.0\textwidth]{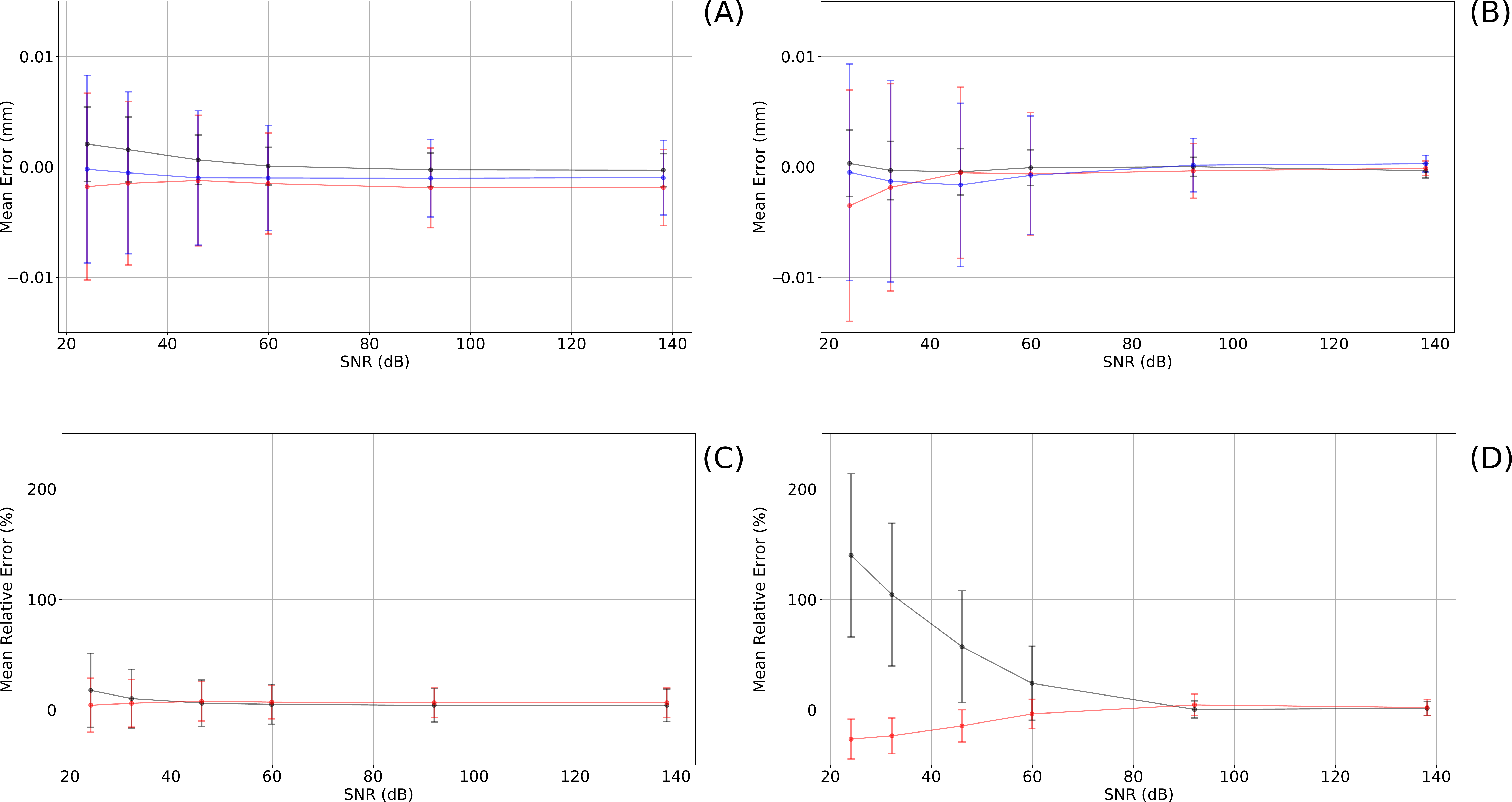}
\caption{(A) $X_{\text{cen}}$, $Y_{\text{cen}}$ and radius (red, blue and black dots, respectively) errors behaviour in presence of noise for the MLP model. (B) $\varepsilon_r$ and $\sigma$ (black and red dots, respectively) errors behaviour in presence of noise for the MLP model. (C) $X_{\text{cen}}$, $Y_{\text{cen}}$ and radius (red, blue and black dots, respectively) errors behaviour in presence of noise for the CNN model. (D) $\varepsilon_r$ and $\sigma$ (black and red dots, respectively) errors behaviour in presence of noise for the CNN model.}\label{Fig3}
\end{figure}

In Fig. \ref{Fig3}, the robustness of the models is tested for each parameter with different Signal-to-Noise Ratios (SNRs). This evaluation was performed by corrupting the test set data with Gaussian noise. In general, MLP networks were more robust than CNNs since with SNRs lower than 50 dB lower errors were obtained. Particularly, the geometric properties estimation seems to have similar behaviours (see Fig. \ref{Fig3} (A) and (B)). The worst performance was observed for the conductivity estimated by the CNNs for noise levels lower than 50 dB, in this case relative errors of 100 \% were obtained. However, for SNR greater than 60 dB the CNNs give always better estimates of all the parameters. \\

\begin{figure}[h]
\centering
\includegraphics[width=1.0\textwidth]{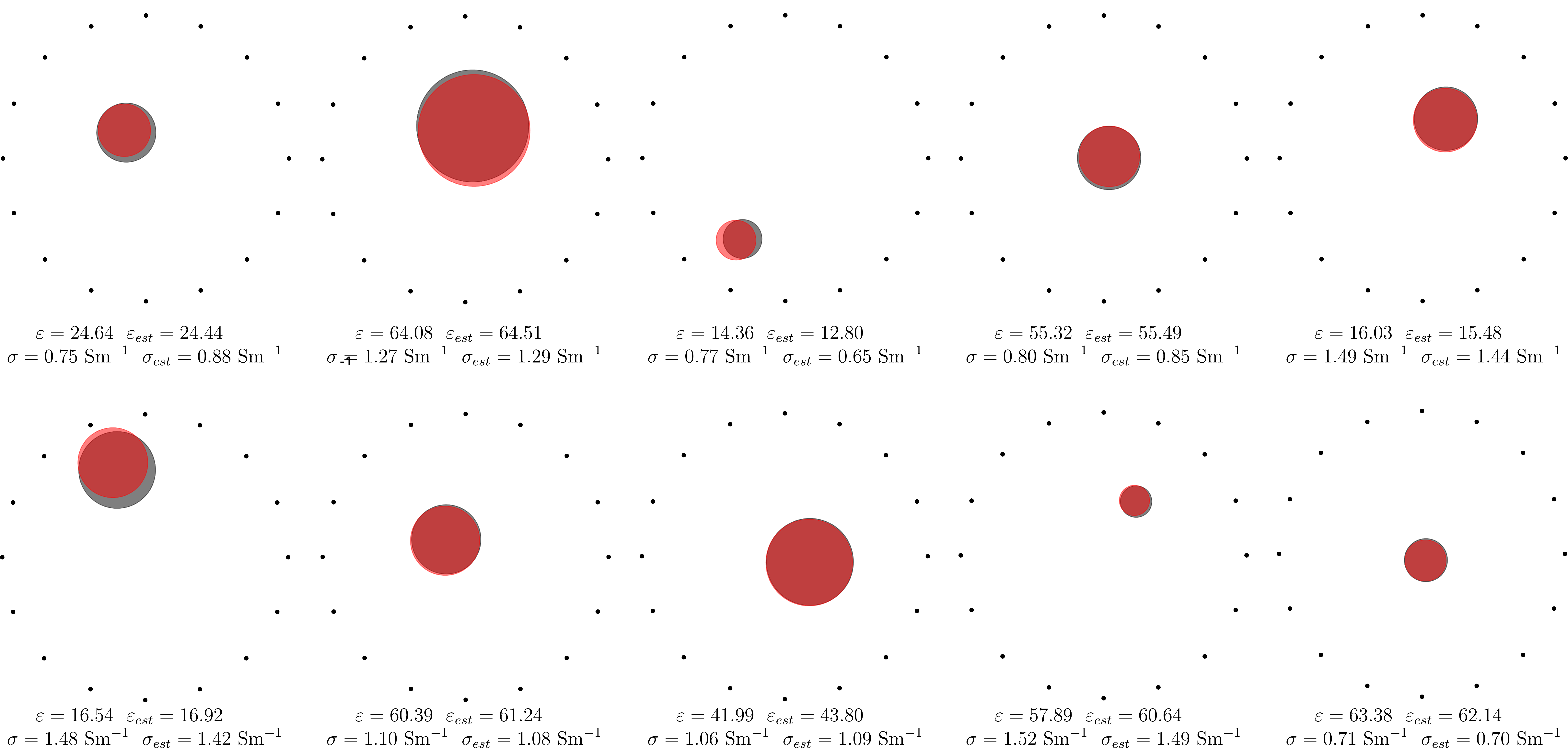}
\caption{Reconstruction examples. The position of the antennas is shown with black dots. The gray circle represents the true cylinder cross-section, the red circle represents the CNN reconstruction. The values of $\varepsilon_r$ and $\sigma$ are also shown for the actual values and the estimated ones (subindex ``est'').}\label{Fig4}
\end{figure}

Some reconstruction examples (with noise free data) for the cylinder dielectric properties, radius and location using the CNN model described in Tab. \ref{TableIII} are shown in Fig. \ref{Fig4}.

\subsection{Prediction of CNN: homogeneous and heterogeneous samples}
In this section we study the behaviour of the previously obtained CNN for the prediction of a hypothetical experimental situation (with SNR greater than 60 dB): the cylinder to be measured should not be in contact with the coupling media. In order to fulfill this condition the cylinder should be put into a sample holder. Two cases were evaluated: homogeneous cylinders and cylinders with cylindrical inclusions (heterogeneous) both into sample holders. We remark that the CNNs were trained with homogeneous cylinders (without holder), consequently, the prediction are effective (or equivalent) values. 

\subsubsection{Homogeneous cylinder in sample container}
\begin{figure}[ht]
\centering
\includegraphics[width=0.7\textwidth]{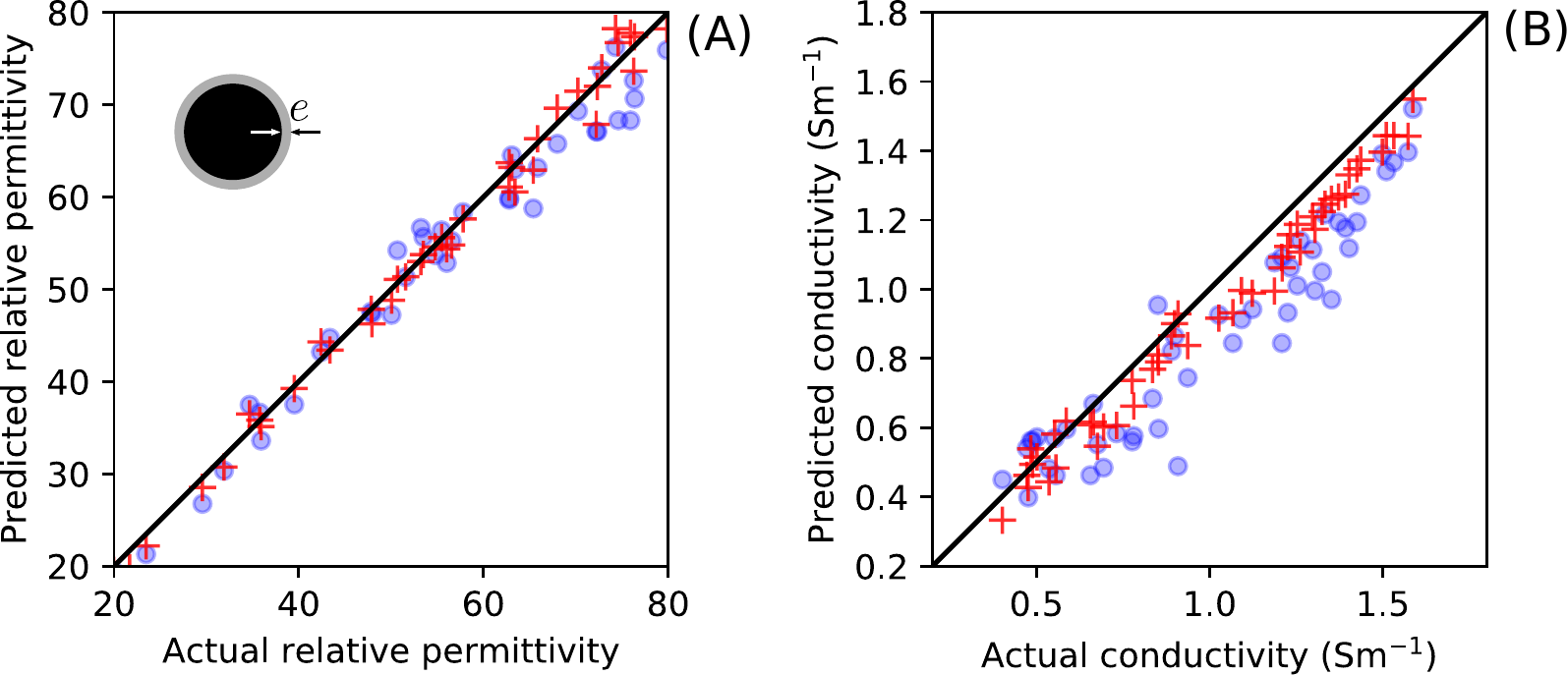}
\caption{Effective relative permittivity (A) and conductivity (B) obtained by CNNs. The circles and crosses indicate the 9 mm and 18 mm radii, respectively. The inset shows the sample in the holder with thickness $e=$ 0.5 mm.}\label{Fig5}
\end{figure}

We considered two size of acrylic sample holders with radii 9 mm and 18 mm, respectively. The thickness was $e=$ 0.5 mm (see inset image in Fig. \ref{Fig5} (A)) and the dielectric properties were $\varepsilon_{r} = 2.0$ and $\sigma\approx 0.0$ for both. Fifty simulations for each container were computed varying the dielectric properties (within the range presented in Table \ref{TableI}) and the position (within the investigation domain) uniformly. The errors in the estimated properties (MAE, see Eq. \ref{mae2}) are shown in Table \ref{TableIV}. Figure \ref{Fig5} shows the results for the effective dielectric properties predictions. For the relative permittivity the estimations fall on the straight line, then the acrylic container seems to not affect its measurement (Fig. \ref{Fig5} (A)). Conductivity estimates are always lower than the actual values and, as expected, this effect worsens for the smaller container. However it is a systematic error that can be corrected experimentally. 

\begin{table}[h!]
\centering
\caption{Mean (standard deviation) of MAE for the homogeneous cylinders within an acrylic holder (expressed $\times$10$^{-3}$).}
\label{TableIV}
\setlength\tabcolsep{1.5pt}
\begin{tabular}{lccc}
\hline
Cylinder & radius & $X_{\text{cen}}$ & $X_{\text{cen}}$ \\
\hline
Homogeneous (radius 0.9 cm) & -0.21 (1.01) & -0.25 (0.96) & 0.15 (0.90) \\
Homogeneous (radius 1.8 cm) & -0.36 (0.98) & 0.15 (1.06) & 0.64 (1.28) \\
Homogeneous 3D (radius 1.7 cm)* & -1.28 & -0.59 & 0.03 \\
\hline
* Just one model was simulated.
\end{tabular}
\end{table}


%

\subsubsection{Heterogeneous cylinder in sample container}

\begin{figure}[ht]
\centering
\includegraphics[width=0.7\textwidth]{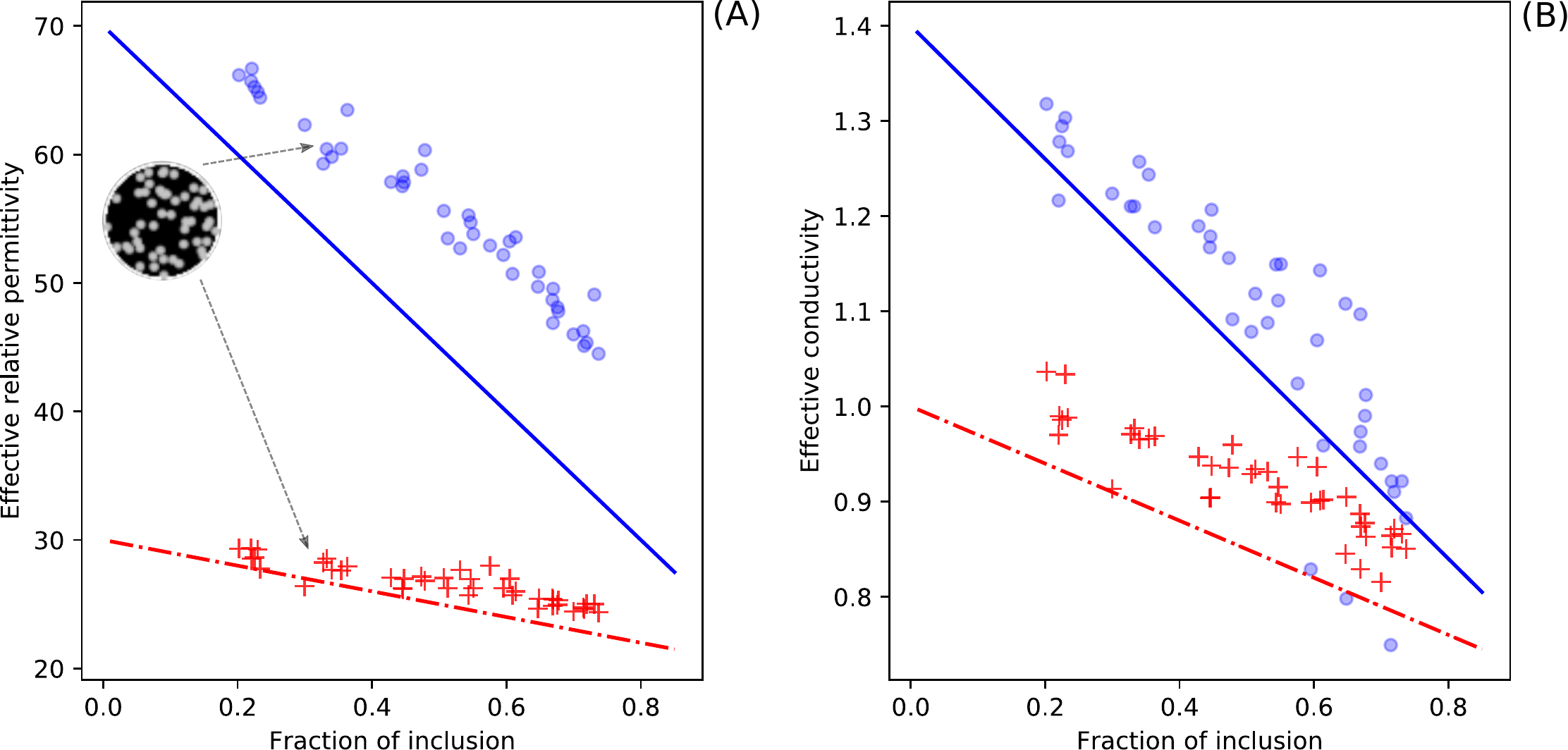}
\caption{Heterogeneous cylinders with variable fraction of inclusion ($\phi_{\text{i}}$) in holder. (A) Relative permittivity. (B) Conductivity. The red crosses and blue circles correspond to case (a) $\varepsilon_{r\text{b}} =$ 70 and $\sigma_{\text{b}}=$ 1.4 Sm$^{-1}$; and case (b) $\varepsilon_{r\text{b}} =$ 30 and $\sigma_{\text{b}}=$ 1.0 Sm$^{-1}$, respectively. The solid lines represent the Wiener models.}\label{Fig6}
\end{figure}

We also tested cylinders with cylindrical inclusions (constant radius, relative permittivity, and conductivity, $r_{\text{i}}= $ 2 mm, $\varepsilon_{r\text{i}} =$ 20, and $\sigma_{\text{i}}=$ 0.7 Sm$^{-1}$, respectively) within an acrylic container (equal to the one of the previous section). The number of inclusions were varied to obtain different fractions of ocuppation and overlap of inclusions was allowed. A sample with fraction of inclusion $\phi_{\text{i}}\approx$ 0.3 is shown in the inset of Fig. \ref{Fig6} (A). Two sets of dielectric properties of the background were evaluated: case (a) $\varepsilon_{r\text{b}} =$ 70 and $\sigma_{\text{b}}=$ 1.4 Sm$^{-1}$; and case (b) $\varepsilon_{r\text{b}} =$ 30 and $\sigma_{\text{b}}=$ 1.0 Sm$^{-1}$. It is important to remark that the radius of the sample and its postition  were also varied (within the range of Table \ref{TableI} and the investigation domain, respectively).\\
There are several effective models (homogenisation of mixtures) to study dielectric samples with inclusions \cite{sihvola1999electromagnetic}. Two broadly used model in such a problems are the Wiener bounds. Particularly, when the electric field is parallel to the direction of the fibers (inclusions of cylindrical shape), the Wiener bound takes the form:
\begin{equation}
\varepsilon^{*}_{\text{eff}}=\varepsilon^{*}_{\text{i}}\phi_{\text{i}}+\varepsilon^{*}_{\text{b}}\left(1-\phi_{\text{i}}\right)
\label{Eq5}
\end{equation}
where $\varepsilon_{j}^{*}=\varepsilon_{r_{j}}-\sigma_{j}/\left(j\omega \epsilon_{0} \right)$ for each component (background and the inclusions, $j=\text{b}$ and $j=\text{i}$, respectively). The effective relative permittivity and conductivity using this bound are also shown in Fig.\ref{Fig6}. It can be seen that both dielectric properties show linear trends but differ from the Wiener bound. The Wiener bound it is not the exact behaviour that such a sample would follow but it seems that it is not so far. For these heterogeneous samples there is also a systematic error that can be corrected experimentally and the fraction of ocuppation could be estimated (if the dielectric properties of inclusions and background are known). Table \ref{TableIV} shows the MAE for the estimated parameters. 

\subsection{Comparison with a 3D simulation}
\begin{figure}[ht]
\centering
\includegraphics[width=0.9\textwidth]{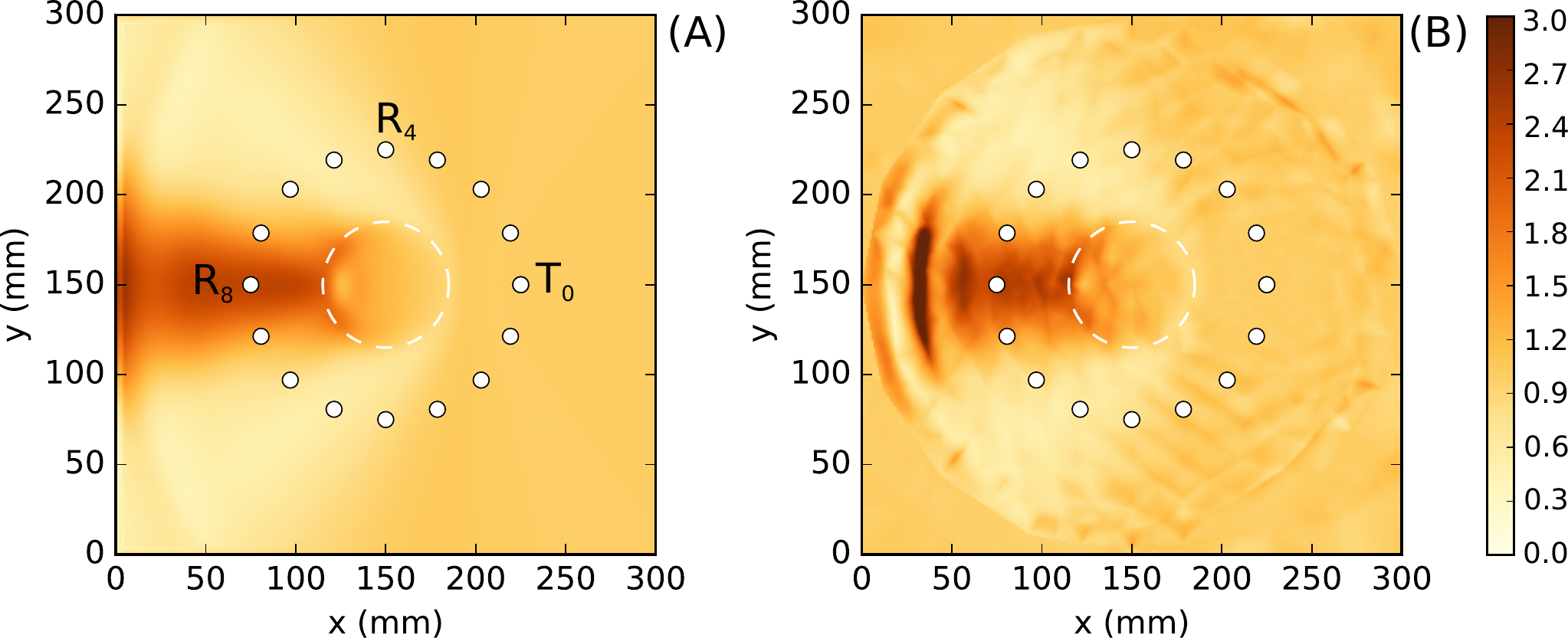}
\caption{Input of the ANN ($\rvert E_{\text{t}} \rvert / \rvert E_{\text{inc}}\rvert $) obtained by simulation of a homogeneous cylinder of 35 mm in radius, 300 mm in length, centered at $(X_{\text{cen}},Y_{\text{cen}})= (0,0)$ m, $\varepsilon_{r}=$ 53.1, and $\sigma=$ 1.43 Sm$^{-1}$.(A) 2D FDTD model and (B) Slice of the 3D FEM model at the center of the monopole antena. White dots and dashed line show the postitions of the antennas and the scatterer cylinder, respectively.}\label{Fig7}
\end{figure}

Models in 3D were computed in order to simulate the whole setup and, consequently, to validate the inverse algorithm using CNN. It is also remarked here that the ANNs were trained with 2D homogeneous cylinders. Homogeneous 3D cylinders (30 cm in length) with fixed postitions at $(X_{\text{cen}},Y_{\text{cen}})= (0,0)$ m were simulated. One transmitter and one receiver was simulated at time. The postition of the receiver was varied from postition 1 to 8 (see Fig. \ref{Fig7}). For example: one simulation is: emitting with the transmitter in position 0 (T$_{\text{0}}$) and receiving with receiver 4 (R$_{\text{4}}$, see Fig. \ref{Fig7} (A)). A total of eigth 3D simulations were computed and the 256 required by the inverse algorithm were obtained using the simetry of the problem. Figure \ref{Fig7} shows the input of the ANN ($\rvert E_{\text{t}} \rvert / \rvert E_{\text{inc}}\rvert $) obtained by the 2D model and the slice of the 3D model (in the center of the active zone of the monopole, that is, the middle of the extended inner connector of the coaxial antenna). Qualitatively, a similar behaviour is observed. In an experimental situation the measured data are the scattering coefficients ($S_{ij}$). Particularly, in the 3D model a lumped port was defined at the coaxial cable and electric fields were converted to current and voltage in order to compute the S-parameters.\\ 

The reconstruction algorithm (using CNN) was tested with one cylinder of 1.7 cm in radius, 30 cm in length, $\varepsilon_{r}= 67.0$ and $\sigma=1.55$ Sm$^{-1}$. The transmission coefficients were only computed ($S_{12}$) and normalized to the $S_{12_{\text{inc}}}$ of the model without cylinder (coupling media only). Finally, $\rvert S_{12} \rvert / \rvert S_{12_{\text{inc}}}\rvert$ data were taken as input of the CNN and the predicted parameters were: 1.587 cm, 0.05 cm, -0.003 cm, 67.6, and 1.58 Sm$^{-1}$, for the radius, $X_{\text{cen}}$, $Y_{\text{cen}}$, relative permittivity, and conductivity relatively. 

\section{Conclusions}
The results presented in this manuscript show that microwave tomography using amplitude--only information to reconstruct a simple geometry as an infinite cylinder is achievable with artificial neural networks. The overall quality of the reconstruction with high signal to noise ration measurements is better with the convolutional neural network model. However, with signal to noise ratios below $\sim$90dB, the position and radius estimations, becomes similar and even worse to those of the multilayer perceptron model (altough less biased). Similar behaviour is observerd with the estimation of the dielectric parameters of the cylinder. In both studied neural network models, the more robust estimations are those of the cylinder position and the radius. Applications for measuring homogeneous and heterogeneous cylinders in acrylic sample holders were proposed. Good results were obtained even with heterogeneous samples with inclusions. Finally, the three dimensional model validate the proposed algorithms which makes this method a potential approach to develop a low cost cylinder imaging setup.

\section*{Acknowledgment}

This work was supported by a grant from the ``Agencia Nacional de Promoción Científica y Tecnológica de Argentina'' (Ref. PICT-2016–2303) and from the ``Universidad Nacional Arturo Jauretche'' (Ref. UNAJ Investiga 2017 80020170100019UJ).

\end{document}